\documentclass{PoS}
\usepackage{subfig}
\usepackage{amsmath}
\usepackage{xspace}
\usepackage{cite}
\def\LIV{\ifmmode {\mathrm{LIV}}\else{\scshape LIV}\fi\xspace}
\def\LI {\ifmmode {\mathrm{LI}}\else{\scshape LI}\fi\xspace}
\def\tot{\ifmmode {\mathrm{tot}}\else{\scshape tot}\fi\xspace}
\def\d  {\ifmmode {\mathrm{d}}\else{\scshape d}\fi\xspace}
\title{The optical depth including Lorentz invariance violation energy threshold shifts}

\ShortTitle{The Optical Depth including LIV}

\author{\speaker{Humberto Mart\'inez-Huerta}$^{\rm ~a}$
		\thanks{The authors acknowledge FAPESP support numbers 2015/15897-1, 2016/24943-0 and 2017/03680-3. The authors also acknowledge the National Laboratory for Scientific Computing (LNCC/MCTI, Brazil) for providing HPC resources of the SDumont supercomputer, which have contributed to the research results reported within this paper (http://sdumont.lncc.br).}
        \\
        E-mail: \email{humbertomh@ifsc.usp.br}}

\author{Rodrigo G. Lang$^{\rm ~a}$
		\\
        E-mail: \email{rodrigo.lang@usp.br}}
       
\author{Vitor de Souza$^{\rm ~a}$
        \\
		E-mail: \email{vitor@ifsc.usp.br} 
		\\ \\
       $^{\rm ~a}$ Instituto de F\'isica de S\~ao Carlos, Universidade de S\~ao Paulo, \\ 
        \ \ \ Av. Trabalhador S\~ao-carlense 400, S\~ao Carlos, Brasil.}

\abstract{

Lorentz invariance violation (LIV) introduced as a generic modification to particle dispersion relations can change the photon energy threshold of pair-production, which modifies the expected gamma-ray flux from astrophysical sources. In this work, we review this phenomenon and explore its consequences through the derived effects in the optical depth. Then, by looking for subluminal LIV signatures in TeV gamma-ray spectra, we present stringent limits to the LIV energy scale at leading order n=1 and 2.  And finally, we present the predicted flux of GZK-photons including LIV, in the astrophysical scenario which best describes UHECR data.
}

\FullConference{International Conference on Black Holes as Cosmic Batteries: UHECRs and Multimessenger Astronomy - BHCB2018\\
		12-15 September, 2018\\
		Foz do Igua\c{c}u, Brasil}

\begin{document}

\section{Introduction}

The precise measurements of cosmic messengers 
from very energetic phenomena in the Universe, have set an unprecedented opportunity to test fundamental physics. 
Among the possibilities, to test and explore the limits of validity of the Lorentz symmetry has been an important motivation for theoretical and experimental research. Moreover, some Lorentz invariance violation (LIV) has been proposed by quantum gravity and effective field theories~\cite{NAMBU, Kostelecky:1988zi, Colladay:1998fq, QG4,QG5, QG1,  ALFARO, Pot, Audren:2013dwa, Bluhm,Calcagni:2016zqv,  Bettoni:2017lxf}.
Therefore, astroparticle signatures of LIV derived phenomena in the photon sector have been searched through the arrival energy time delays, photon splitting, spontaneous photon emission, shifts in the pair production energy threshold and many others effects~\cite{bib:liv:tests:astropart, Coleman:1997xq, AmelinoEllis:1998, Coleman:1998ti, Stecker:2001vb, Stecker:2003pw, Jacobson:2002, Stecker:2004, Ellis2006402, Gunter:2007, Gunter:2008, ALBERT2008253,    Stecker:2009,   XU201672,1674-1137-40-4-045102, ELLIS201350,   Farbairn:2014,  Tavecchio:2015, Biteau:2015, Martinez-Huerta:2016azo, Rubtosov:2017, Martinez-Huerta:2017gna, Lang:2017wpe,  Cologna:2016cws, Mrk501_HESS_flare,  Pfeifer:2018pty,  Ellis:2018lca,Abdalla:2018sxi,Lang:2018yog}.
Following this line of thought, in the present proceeding we review and show some recent results of our LIV signatures searches in cosmic and gamma rays. In Section \ref{sec:LIV}, we present the phenomenological generalization of the LIV-induced modifications to the particle dispersion relation.
Then, we review and explore the scenarios when LIV is considered in the electron-positron production by a high energy photon that interacts with some low energy background light and show that the LIV effect is a shift to the minimum energy that the pair-production process requires, we present this in Section \ref{sec:PP}. The consequence of such effects, in the so called {\it subluminal} LIV-scenarios, is the increase of the optical depth at given ener\-gy regions, which, as we illustrate in Section \ref{sec:OD}, are sensitive to the amount of LIV. The latter predicts that more photon events can be expected in different energy windows than in a Lorentz invariant (LI) scenario. Thus, in Section \ref{sec:TeV}, we present results of a LIV signature search in TeV gamma-rays. Then, in Sec. \ref{sec:GZK}, we show the expected, LI and LIV, EeV photon flux produced by the decay of the secondary cosmic ray flux of neutral pions, the usually named GZK-photons. Finally, in Section~\ref{sec:C}, some conclusions and remarks are presented. 

\section{Lorentz invariance violation}
\label{sec:LIV}

A phenomenological generalization of the LIV-induced modifications to the dispersion relation formalism converges to the introduction of an extra term in the $a$-specie particle dispersion relation, 
which, among other possibilities, can be motivated by the introduction of a not explicitly Lorentz invariant term in the free particle Lagrangian \cite{Coleman:1997xq, Coleman:1998ti} or by some spontaneous Lorentz symmetry breaking \cite{Kostelecky:1988zi}. So that, the corrected dispersion relation becomes
\begin{equation}\label{eq:1}
E^2_{a} - p^2_{a} = m^2_a \pm |\delta_{n,a}|  E_a^{(n+2)},
\end{equation}
where $(E_a,p_a)$ stands for the four-momenta of the $a$-particle type with mass, $m_{a}$. For  simpli\-city,  natural units  are used in this work, unless a different one is explicitly given. 
The ($\pm$) sign, cha\-racterize the so called {\it superluminal} (+) and {\it subluminal} (-) dominant phenomena due to $\delta_{n,a}$, which is the Lorentz invariance violation parameter, where $n$ express the leading order of the correction. $\delta_{n,a}$ is frequently considered to be inversely proportional to some LIV energy scale $\rm M$,
to su\-ppress higher-dimension operators with some coefficients given by the underlying theory, and it is also common to associate ${\rm M}$ with the Plank energy scale ($\sim$~$10^{19}$~GeV). However, without loss of generality, the LIV term can be named $E_{\LIV}^{(n)}= 1/|\delta_{n}|^{1/n}$ for $n>0$, as we do in Section~\ref{sec:TeV}. 

It has been shown in previous works that processes of photo production can lead to new physics when LIV is considered through Eq.~\ref{eq:1}, and there is some shift of the minimum energy that these processes need to be kinematically allowed \cite{Stecker:2001vb, Stecker:2003pw, Gunter:2007, Gunter:2008, Martinez-Huerta:2016azo}, which it is discussed in the next Section. 

\subsection{Pair production energy threshold shifts}
\label{sec:PP}

In the particular scenario where the interaction of very energetic particles, such as cosmic and gamma rays, propagating through the background light of the Universe, LIV can lead to measurable signatures in the observed cosmic particle spectra, due to changes in the expected effects of relevant processes of photon-particle production, such as photo pion-production and pair-production processes. 
Hereafter, we focus in the latter due to its strong impact as an energy lose mechanism in the propagation of very high energy photons. Although, other processes such as the Compton sca\-tte\-ring can also lead to LIV signatures that modify the observed photon energy spectrum, it is noted in Ref.~\cite{Abdalla:2018sxi}, that its effects are small in comparison with those expected by photo pair production. 

The derived physics from Eq.~\ref{eq:1} leads to shifts at the minimum energy that these processes needs to be kinematically allowed, which can change the expected photon flux from distant sources, as we explore in the next Sections. In the case of the pair-production process\footnote{Sub-index $b$ indicates the low energy photons from the background light with energy $\epsilon$.}, 
$\gamma \ \ \gamma_{b} \longrightarrow \ e^+ e^-,$
if LIV effects are considered for photons and leptons, Eq.~\ref{eq:1} takes the form
\begin{equation}\label{eq:DR}
	E_{\gamma}^2 - p_\gamma^2 = \delta_n E_{\gamma}^{n+2},
	\ \  \ {\rm and}   \ \ \ \ \
	E^2_{\pm} - p_{\pm}^2 = m_e^2 + \delta_{\pm,n} E_\pm^{n+2},
\end{equation}
where sub-indices $\gamma$, $+$ and $-$, denote photons, positron and electron species respectively.
Due to the very high energy of the gamma-rays that we are considering, hereafter, we take $E_{\gamma} \gg m_e, \epsilon$, thus, the kinematics of the process for a head-on collision, with inelasticity $K$, ($E_{+}= K (E_\gamma + \epsilon)$), satisfies the following expression \cite{Martinez-Huerta:2016azo},
\begin{equation}\label{eq:TH1}
\epsilon = \frac{m_e^2}{4E_{\gamma} K(1-K)} - \frac{1}{4} \delta^{tot}_n \ E_{\gamma}^{n+1} \ ;  
\ \ \ \ \ \ \ \ \  
\delta^{\rm tot}_n =  \delta_{\gamma,n} - \delta_{+,n}K^{n+1} - \delta_{-,n}(1-K)^{n+1}.
\end{equation}
Notice that  $\delta_n^{tot}$ is a linear combination of the LIV contributions from the different particle species. %
%
Also note that, if LIV is such that, $\delta_n=\delta_{\gamma,n}=\delta_{\pm,n}$, for the scenario with $K=1/2$,  
\begin{equation}
  \delta^{\rm tot}_n=~(1~-~\frac{1}{2^n}) \delta_n.  
\end{equation}
On the other hand, when  LIV is considered only (or dominated by) the photon sector, i.e.  $\delta_{\pm}=0$ (or $\delta_{\gamma} \gg \delta_{\pm}$), then, $\delta_{n}^{\rm tot} = \delta_{\gamma,n}$ (or $\approx \delta_{\gamma,n}$). Therefore, there is only a factor of  $(1-\frac{1}{2^n})$ between this two set of scenarios. So that, for simplicity and as in previous works, these second scenarios are considered in Sections \ref{sec:TeV} and \ref{sec:GZK}, which also allows the particular case of LIV at $n=0$. 

\begin{figure}[t!]
    \centering
    \includegraphics[width=.80\linewidth]{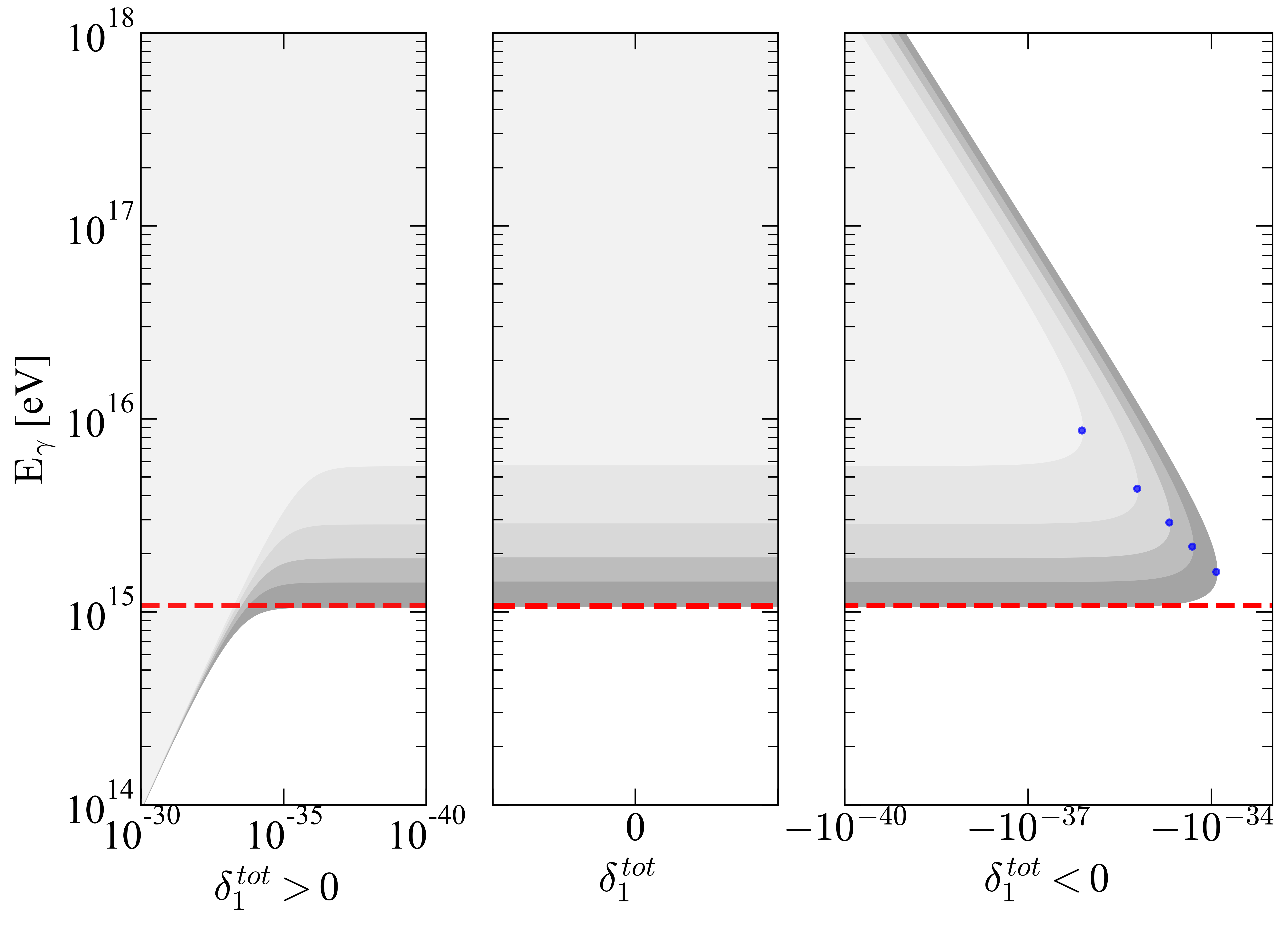} 
    \caption{\small The three different scenarios when LIV effects are considered in the particular case of, $\gamma \ \gamma_{b}\rightarrow e^+ \  e^-$ and LIV leading order $n=1$. Grey regions show the pair-production allowed configuration at background temperatures of $0.5, 1, 1.5, 2$ and $2.7$ K, from lighter to darker greys, where each shaded area includes the previous ones. Similar behaviour can be expected for different temperatures and leading orders $n$, but different energies. 
    At the central panel, the LI scenario, $\delta^{\rm tot}_n = 0$, thus, 
    $E_{\gamma,th}^{\LIV}=E_{\gamma,th}^{\LI}$.
    In the left panel, there are the {\it superluminal} scenarios, $\delta^{\rm tot}_n > 0$, 
    where the LIV energy threshold is $E_{\gamma,th}^{\LIV}\leq E_{\gamma,th}^{\LI}$.
    At the right panel, there are the {\it subluminal} scenarios, $\delta^{\rm tot}_n < 0$, 
    where energy threshold shift is such that $E_{\gamma,1,th1}^{\LIV} \geq E_{\gamma,th}^{\LI} $ and the pair production is constrained to a region $E_{\gamma,1,th1}^{\LIV} \geq E_{\gamma} \leq E_{\gamma,1,th2}^{\LIV}$, which gets shorter as the LIV effect increases, until it reaches a critical point (in blue), where no process is allowed at all. Contrary to the previous scenario, more gamma-rays can be expected. The red dotted line is the $E_{\gamma,th}^{\LI}$ at 2.7~K for comparison.}
    \label{fig:scenarios}%
\end{figure}

Solutions for one of the photon energies in Eq. \ref{eq:TH1} are the photon energy thresholds, $E_{\gamma,th}$ or $\epsilon_{th}$, that turns on the process. Solving for $E_{\gamma,th}$, Eq. \ref{eq:TH1} is a polynomial equation of order $n$ that has three main scenarios depending on $\delta^{\rm tot}_n$. a) If $\delta^{\rm tot}_n = 0$, the standard LI-threshold is recovered, which correspond to the central panel in Fig.\ref{fig:scenarios}, where pair-production is allowed for any $E_{\gamma}$ once the gamma-ray energy reaches the grey regions. As an example, in the Figure \ref{fig:scenarios} there are different areas marked for different background temperatures, $0.5, 1, 1.5, 2$ and $2.7$ K, from lighter to darker grey, where each shaded area includes the previous ones.  
However, as pointed in Ref. \cite{Martinez-Huerta:2016azo}, if the LIV parameter is not zero, there are two more scenarios depending on the sign of $\delta_{n}^{\rm tot}$. 
b) If $\delta_{n}^{\rm tot} > 0$,  then $E_{\gamma,th}^{\LIV}\leq E_{\gamma,th}^{\LI}$, that is, there is a threshold shift to lower energies as displayed in the left panel of Fig.~\ref{fig:scenarios}. For comparison, the red dotted line show the $E_{\gamma,th}^{\LI}$, at $2.7$ K.
c) If $\delta_{n}^{\rm tot} < 0$, the so called "recovery scenarios", there is a threshold shift to higher energies and the process is allowed only in the region $E_{\gamma,th_1}^{\LIV}<E_{\gamma}<E_{\gamma,th_2}^{\LIV}$, as shown in the right panel of  Fig.~\ref{fig:scenarios}. Moreover, if $|\delta_{n}^{\rm tot}|>|\delta_{n}^{cr}|$, the LIV effect is so strong that no pair-production is allowed at all. These critical points are indicated with the blue points in Fig. \ref{fig:scenarios} and given by
\begin{equation}\label{eq:CR}
    E_\gamma^{cr} (\epsilon; n) = \frac{n+2}{n+1} \frac{m_e^2}{4K (1-K)}\frac{1}{\epsilon} \ , \ \ \ \ \
    \delta_{n}^{cr} (\epsilon; n) = \frac{(n+1)^{n+1}}{(n+2)^{n+2}}
    \left(\frac{4K (1-K)}{m_e^2} \right)^{n+1} 4 \epsilon^{n+2}.
\end{equation}
In the next Section, we explore the implications of gamma-rays interacting with the different dominant background lights in different energy regions of interest.

\section{The Optical Depth including LIV}
\label{sec:OD}

As pointed out, very high gamma-rays that propagate from distant sources suffer significant attenuation due to pair-production, which constrains how far in the universe we, on Earth, can expect photons without being absorbed, which is the optical depth \cite{DeAngelis:2013jna}.  Therefore, the effect of shifting the energy threshold of pair-production leads to a change in the LI-expected gamma-ray flux. 

It has been shown in previous works (see for instance Refs. \cite{Biteau:2015, Cologna:2016cws, Mrk501_HESS_flare,  Lang:2017wpe, Lang:2018yog}), that the optical depth including LIV effects can be obtained by
\begin{equation}\label{eq:tau}
\small{
\tau_{\gamma}(z,\theta,\eta_b, E_{\gamma};n,\delta_{n}^{\tot}) = \int_{0}^{z} \d z \frac{c}{H_{0} (1+z) 
h(z) } \ 
\int_{-1}^{1} \d(\cos\theta) \frac{1-\cos{\theta}}{2}  \ 
\int_{\epsilon_{th}}^{\infty} \d\epsilon \ \eta_{b} (\epsilon,z) \ \sigma(E_\gamma,\epsilon,z),
}
\end{equation}
where $H_{0} = 70$ km s$^{-1}$ Mpc$^{-1}$ is the Hubble constant,  $c$ is the speed of light in vacuum, $h(z) = \sqrt{\Omega_{\Lambda} + \Omega_{M} (1+z)^{3}}$ is the distance element in a expanding universe with, $\Omega_{\Lambda} = 0.7$ and $\Omega_{M} = 0.3$,  $\sigma$ is the cross-section of the pair-production process, $\theta = [-\pi,+\pi]$ is the angle between particles, $\eta_{b}$ is the background photon density and $\epsilon_{th}$ is the background photon energy  threshold as given by Equation~\ref{eq:TH1}. 
Due to the nature of the background light in the universe, there are dominant $\eta_{b}$ in different energy regions that need to be addressed to properly estimate $\tau_{\gamma}$.  The ones we consider are the extragalactic background light (EBL), for $E_\gamma < 10^{14.5}$ eV, the cosmic background microwave radiation (CMB) for $10^{14.5} \ \mathrm{eV} < E_\gamma < 10^{19}$ eV and the radio background (RB) for $E_\gamma > 10^{19}$ eV.

\begin{figure}[t]
    \centering
    \subfloat[]{{\includegraphics[width=.46\linewidth]{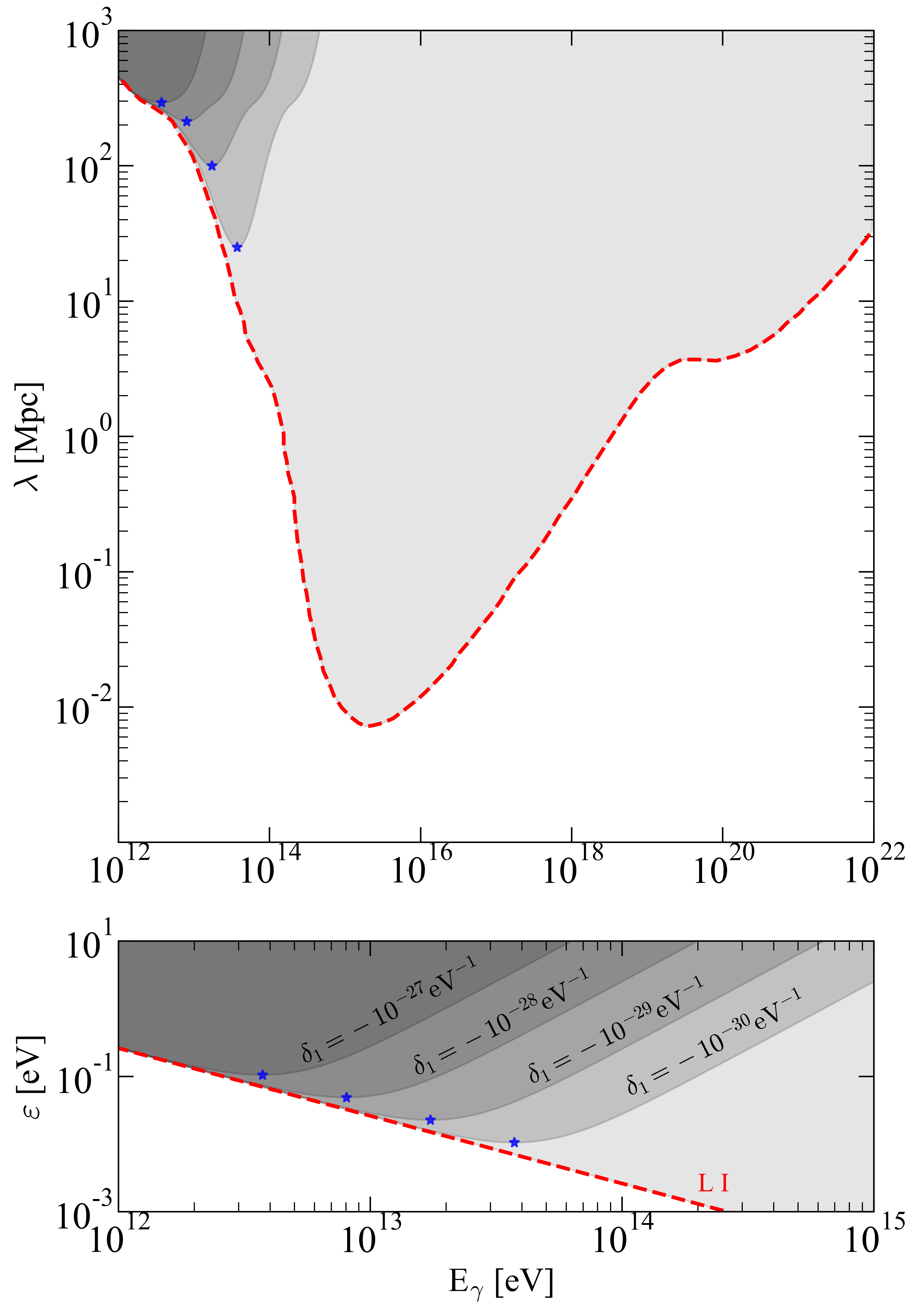} }}%
    \qquad
    \subfloat[]{{\includegraphics[width=.46\linewidth]{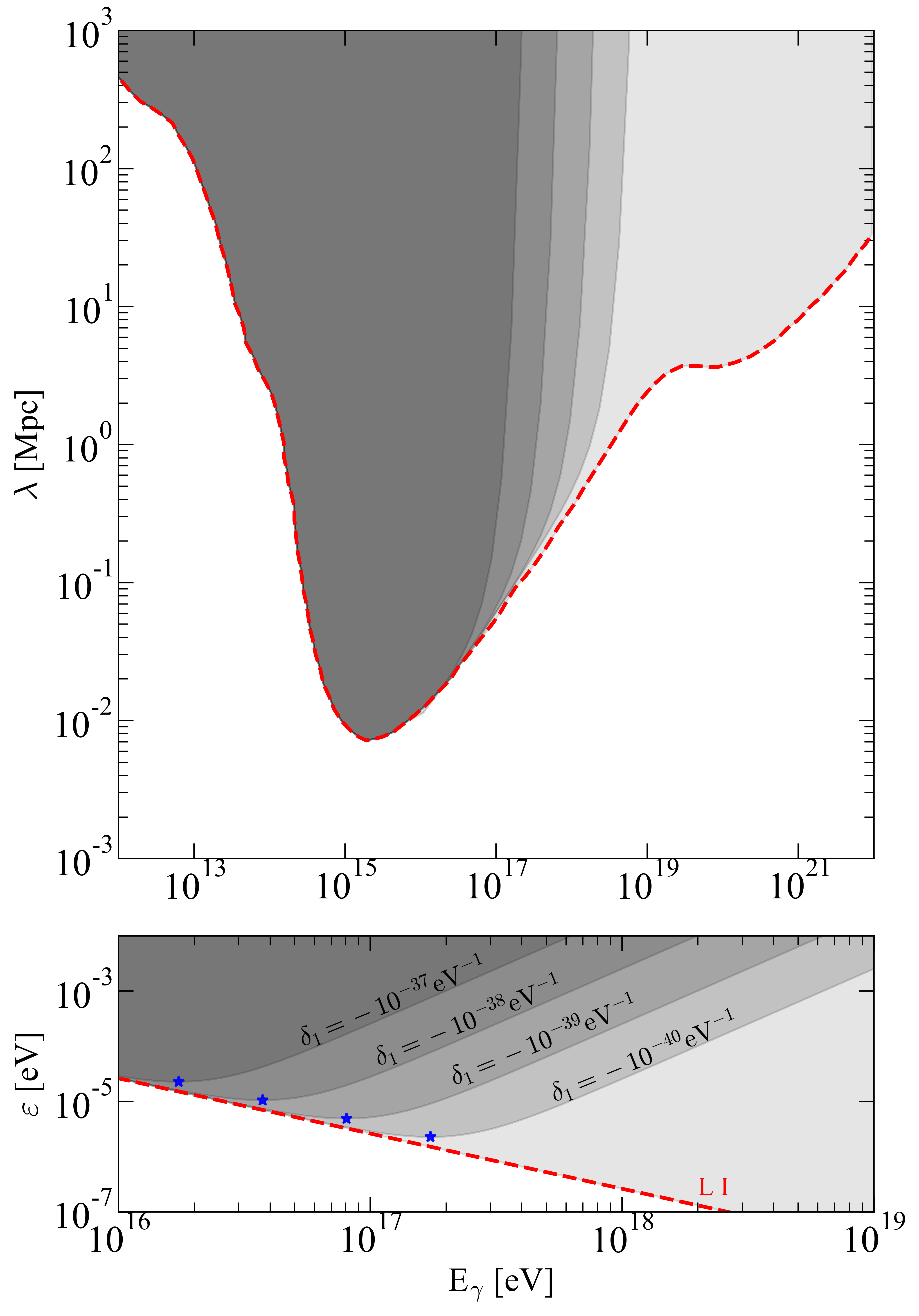} }}%
    \caption{ In the top panels, the photon mean-free path as function of $E_\gamma$ at EBL (a) and CMB (b) regions.
    The red dotted line is the LI expected behaviour ($\delta_{1} = 0$). The shaded regions are the none transparent universe as a function of the gamma-ray energy, $E_{\gamma}$, in the different scenarios. From the darker to lighter grey: (a) $\delta_{1}= ( -10^{-27},-10^{-28},-10^{-29},-10^{-30},\ 0)$ eV$^{-1}$ and (b) $\delta_{1}= (-10^{-37},-10^{-38},-10^{-39},-10^{-40},\ 0)$ eV$^{-1}$. Each shaded area includes the previous one.  In the bottom panels, allowed regions for the pair production process $\gamma\gamma_{b}\rightarrow e^+ e^-$ in the same scenarios that the top panels. Blue stars show the critical values in Eq. \ref{eq:CR}.}
    \label{fig:OD}%
\end{figure}

The resulting mean-free path, $\lambda =(c z)/({\rm H_0} \tau_{\gamma})$, using Eq.~\ref{eq:tau}, at the EBL\footnote{In the calculations presented here, the Franceschini {\it et al.} EBL model is used \cite{Franceschini}.} and CMB regions, are presented in the left and right top panels of Fig.~\ref{fig:OD}. For simplicity, we consider that the LIV in the process is dominated by the {\it subluminal} photon sector with leading order $n=1$, that is $\delta^{\rm tot}_{1} = \delta_{\gamma,1} := \delta_{1}$, where $\delta_{1}<0$. Once again, the red dotted line in the Figure is the LI expected behaviour and the shaded regions are the none transparent universe as a function of the gamma-ray energy, $E_{\gamma}$, in the different scenarios. 
The LIV scenarios for (a) $\delta_{1}= ( -10^{-27},-10^{-28},-10^{-29},-10^{-30},\ 0)$ eV$^{-1}$ and (b) $\delta_{1}= (-10^{-37},-10^{-38},-10^{-39},-10^{-40},\ 0)$ eV$^{-1}$ are shown in the grey areas from darker to lighter, where each area includes the previous one. As can be seen, when LIV is considered, the opacity of the universe change in such a way that allows more photons to arrive from further distances and sources, which have the potential to be measured, which we explore in the next Section.  In the bottom of the Figure, there are the energy photon space for each scenario that shows the decrease of the allowed configurations in each different scenario in the top panels. Similar results are expected for $n=0$ and $2$ \cite{Lang:2017wpe}. 

\subsection{TeV gamma-rays absorption including LIV}
\label{sec:TeV}

\begin{figure}[t]
  \centering
    \subfloat[]{{\includegraphics[width=.47\linewidth]{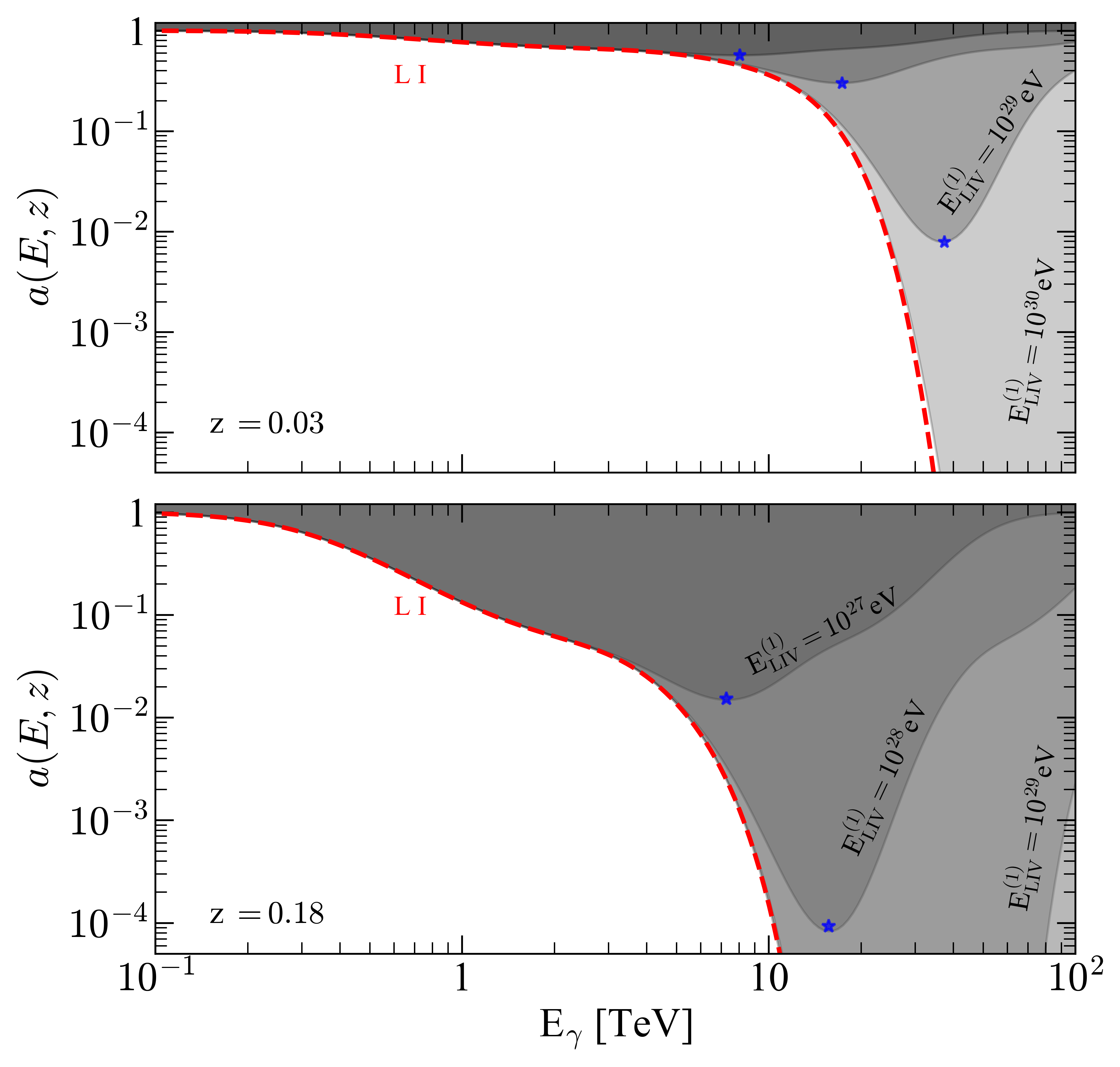} }}%
    \qquad
    \subfloat[]{{\includegraphics[width=.42\linewidth]{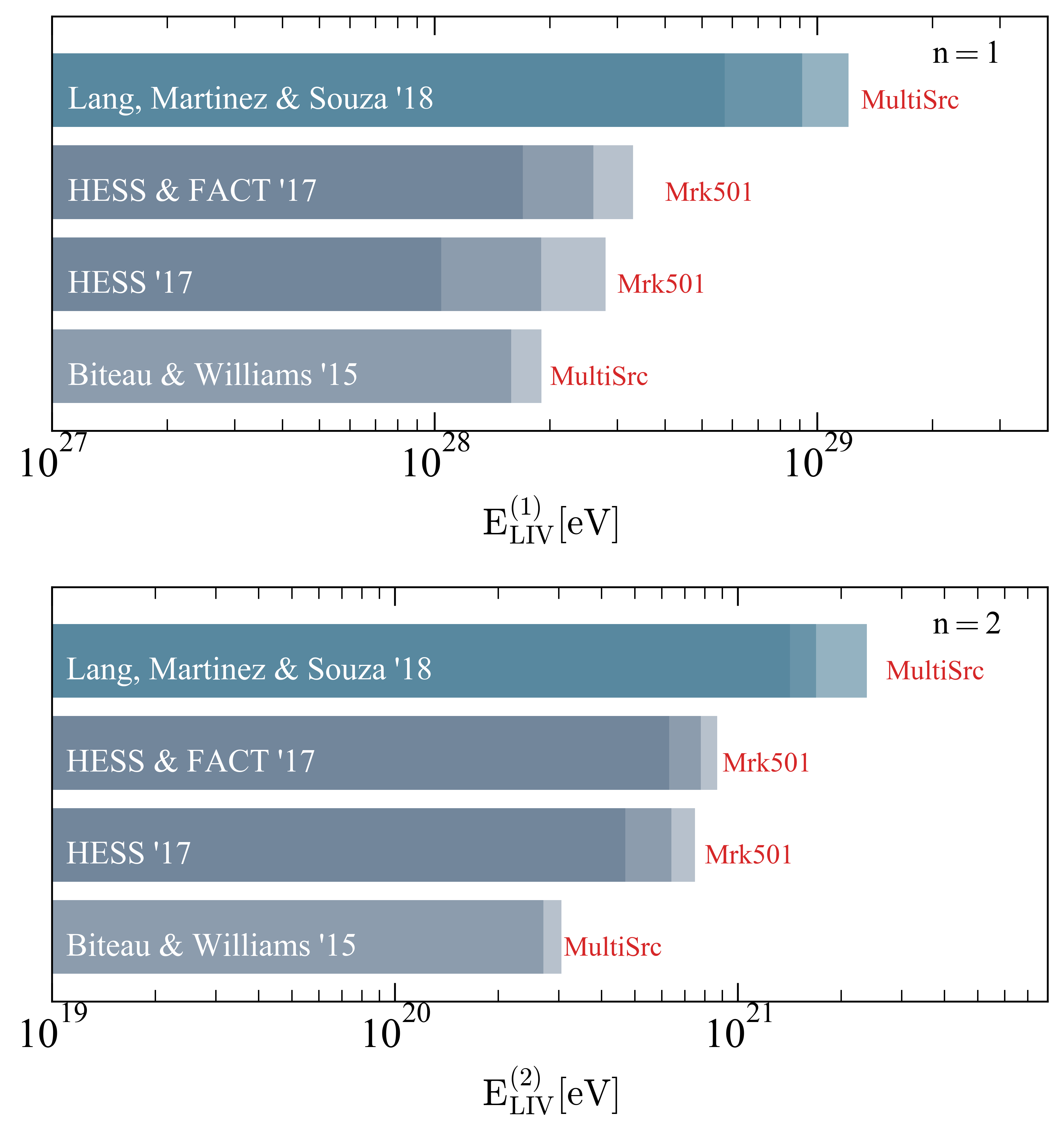} }}%
  \caption{\small (a) Gamma-ray attenuation for LI and LIV cases, with $E_{\mathrm{LIV}}^{(1)} = (10^{27}, 10^{28}, 10^{29}, 10^{30})$ eV. Blue stars show the critical values given by Eq. \ref{eq:CR}. (b) Comparison of limits imposed on the LIV energy scale. Top panel for $n=1$ and bottom panel for $n=2$. Shades of blues correspond to 2, 3 an 5 $\sigma$ CL.}
  \label{fig:TeV}
\end{figure}

In the region of interest with the EBL dominant background, TeV gamma-rays are the ones that start to suffer energy loss due pair-production, so, LIV signatures can be expected in the shape of the attenuated spectra of TeV sources. 
The left panel of Figure \ref{fig:TeV} shows the behaviour of the gamma-ray attenuation, 
\begin{equation}
  a(E_{\gamma},z; \delta_{n}^{tot},n)= \exp{-\tau},  
\end{equation}
at z = 0.03 and 0.18,  for the scenarios where {\small $ \delta_{1} = (0, -10^{-27}, -10^{-28}, -10^{-29}, -10^{-30})$} eV$^{-1} $ and $\tau$ is given by Eq. \ref{eq:tau}. %
In the red dotted line there is the LI scenario, $\delta_{1} = 0$, which has a hard drop in the attenuation curve. However, when LIV is considered, a {\it recovery} in the photon flux can be expected, with an inflection point given by Eq. \ref{eq:CR}, i.e. $\epsilon^{cr} (n, \delta_{n}^{tot}) = [  \frac{1}{4} \frac{(n+2)^{(n+2)}}{(n+1)^{(n+1)}}  
(\frac{m_e^2}{4K(1-K)})^{(n+1)}   \delta_{n}^{tot} ]^{1/(n+2)}$ 
and 
$E_{\gamma}^{cr}(n,\delta_{n}^{tot}) = [ \frac{4}{n+1} \frac{m_e^2}{4K(1-K)} \frac{1}{\delta_{n}^{tot}}]^{1/(n+2)} $. Shaded areas show where photon events are constrained by the EBL absorption. Each area is inclusive from darker to lighter grey. 

In Reference \cite{Lang:2018yog}, we proposed a new analysis procedure to search LIV signatures in this channel with the most updated TeV gamma-ray dataset. In there, we look at 111 measured energy spectra from 38 sources, and found that only 18 measured spectra from 6 sources are expected to have significant contribution to restrict the LIV energy scale beyond the current limits to the $E_{LIV}^{(n)}$. We conclude that the dataset is best described by the LI assumption and we set stringent limits to the LIV energy scales at 2, 3 and 5 $\sigma$ (CL): 
E$^{(1)}_{\LIV} = \{12.08, \ 9.14, \ 5.73\}\times 10^{28}$ eV and 
E$^{(2)}_{\LIV}~=~\{2.38, \ 1.69, \ 1.42\}\times 10^{21}$ eV. 
In addition, it was shown in Ref. \cite{Lang:2018yog}, that the results are robust under poor knowledge of the EBL, large uncertainties in the intrinsic energy spectra functional form, energy resolution, selection of spectra, and energy bin selection used in the calculation of the intrinsic energy spectra. 
The right panel in Fig. \ref{fig:TeV} compares these results with previous limits established by similar LIV signature searches reported in the References \cite{Mrk501_HESS_flare,  Cologna:2016cws, Biteau:2015}.

\subsection{GZK photon flux including LIV}
\label{sec:GZK}

Kenneth Greisen, Vadim Kuzmin and Georgiy Zatsepin (GZK) showed that UHECR that pro\-pagates from further sources have a probability to interact with the CMB and generate a secondary flux of pions via the $\Delta$ resonance, which constrains the propagation distance of UHECR~\cite{GZK-GK, GZK-Z}. Neutral pions decay into a couple of photons most of the times, leading to an expected flux of photons, which are usually named GZK-photons. 

\begin{figure}[t!]
    \centering
    \includegraphics[width=.70\linewidth]{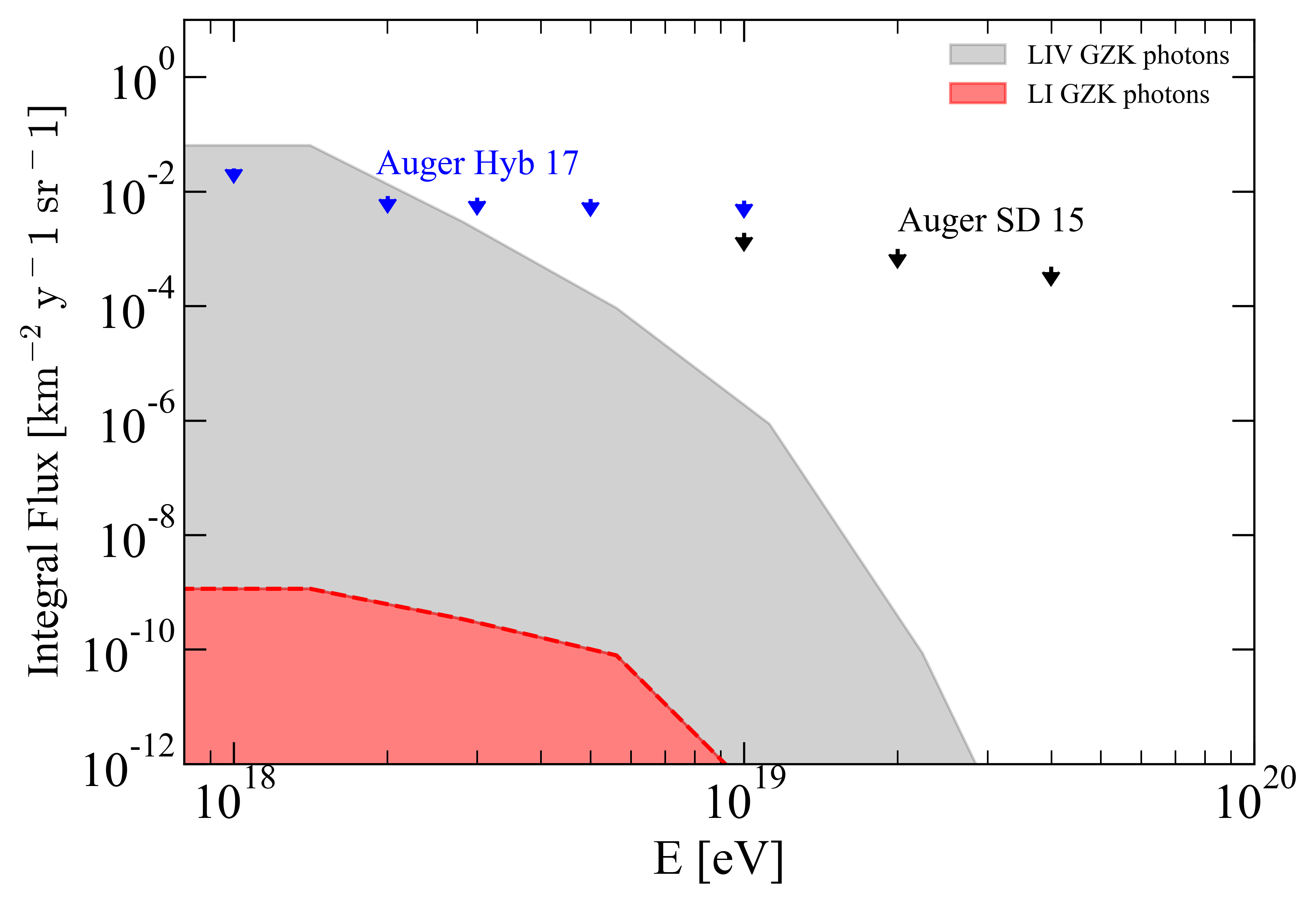} 
    \caption{\small Integral flux of GZK-photons as a function of the photon energy, $E_{\gamma}$, with and without the LIV effects in the model that was shown to best describe the energy spectrum, composition, and arrival direction of UHECR. The arrows correspond to the upper limits to the integrated photon flux from the Pierre Auger Observatory (Auger SD (2015) \cite{UpperLimitsSD} and Auger Hyb (2017)\cite{UpperLimitsAuger}). }%
    \label{fig:GZK}%
\end{figure}

As we commented in the previous Sections, the introduction of some LIV can shift the pair-production energy threshold, which in the {\it subluminal} case leads to an increase of the expected photon flux. Thus, in Ref. \cite{Lang:2017wpe} we proposed a search for LIV signatures at the CMB dominant background region, by computing the GZK-photon flux on Earth considering, for the first time in the literature, several UHECR injection models and source distribution models. Moreover,  Reference~\cite{Unger:2015laa} shows that the best combination of the two type of models that was shown to best describe the energy spectrum, composition, and arrival direction of UHECR corresponds to the one with a source distribution model that follow a GRB rate evolution proportional to $(1 + 11z)/[1 + (z/3)^{0.5}]$ \cite{R5,Kotera:2010yn}, and with an injection model that considers a power law energy spectrum at the source with a rigidity cutoff, given by 
\begin{equation}
\centering
\frac{dN}{dE_{s}} = \begin{cases}
E_{s}^{-\Gamma} \ \ \ \ \ \ \ \ \ \ \ \ \ \ \  \text{, for } R_{s} < R_{cut} \\
E_{s}^{-\Gamma} e^{1 - R_{s}/R_{cut}} \ \text{, for } R_{s} \ge R_{cut}
\end{cases} ,
\end{equation}
where $\Gamma = 1.25$ and $\log_{10} (R_{cut}/V) = 18.5$, and also consider five different species of primary cosmic ray nuclei, with fractions: 
$fH = 0.365$, $fHe=0.309$, $fN=0.121$, $fSi=0.1066$ and $fFe~=~0.098$. 
In Ref. \cite{Lang:2017wpe} other $\Gamma$, $R_{cut}/V$ and fraction values were also used, including a pure proton scenarios and a Si one, combined with other 4 different source distribution models.
The integrated GZK photon flux in the different scenarios were computed with the modified CRPropa3/EleCa codes \cite{bib:crpropa:3}. As expected, the LIV effect is the increase in the predicted GZK-photon flux, which can be appreciated in Fig. \ref{fig:GZK}. The expected region for the GZK-photons in the LI-scenario is the bottom red shaded area and the grey top region shows the one for the GZK-photons including LIV, where the lower limit is the LI-scenario and the top one, where the LIV effect is maximum, $\delta_{n}\rightarrow -\infty$. 
Comparing this results with the new upper limits of the photon flux obtained by the Pierre Auger Observatory (Auger SD (2015) \cite{UpperLimitsSD} and Auger Hyb (2017)\cite{UpperLimitsAuger}), in the black and blue arrows, the LIV-scenarios with $\delta_{0} \sim -10^{-20}$, $\delta_{1} \sim -10^{-38} \mathrm{eV}^{-1}$ and $\delta_{2} \sim -10^{-56} \mathrm{eV}^{-2}$ are compromised. Although, these limits are several orders of magnitude more restrictive than those presented in the previous Section, the comparison is not straightforward due to the different systematics of the measurements and the photon energy region. 

\section{Conclusions}
\label{sec:C}

In this proceeding we have reviewed and explored the effects of the phenomenological ge\-ne\-ra\-lization of the LIV-induced modifications to the dispersion relation via Eq.\ref{eq:1}, that leads, in the subluminal scenario, to an increase in the number of photons that can be expected in the TeV and EeV energy regions, when compared with the LI case, due to a shift in energy threshold of the pair-production process. 

We have presented and discussed the optical depth including significant LIV effects in both regions of interest. In the TeV gamma-ray sector, we have explored the LIV signatures in the TeV spectra and presented stringent limits to the LIV energy scale in the subluminal and photon sector, at leading order $n=1$ and $2$. At the EeV energy region, we have presented the predicted GZK-photon flux including LIV effects for the scenario that is most compatible with the most updated data of UHECR.

Astroparticle physics has reached the status of precision science due to the development and construction of new observatories, operating innovative technologies and the detection of large numbers of events and sources, which set an unprecedented opportunity to test fundamental physics, such as effects of some Lorentz invariance violation as we have addressed in this pro\-cee\-ding. Moreover, updates and new studies can be expected with the advent of new and better data from the cosmic messengers. 


\providecommand{\newblock}{}

\end{document}